\documentstyle[aps,twocolumn,epsf]{revtex}

\newcommand{\hbt}{\mbox{\boldmath{$\widehat{t}$}}}
\newcommand{\bnu}{\mbox{\boldmath{$\nu$}}}

\newcommand{\bbs}{\mbox{\boldmath{$s$}}}
\newcommand{\bn}{\mbox{\boldmath{$n$}}}

\begin{document}

\title{Finite size effects and error-free communication in Gaussian channels}
\author{Ido~Kanter$^{*}$ and David~Saad$^{\#}$}
 \address{$^{*}$ Minerva Center and Department of Physics, Bar-Ilan
University, Ramat-Gan 52900, Israel. \\ $^{\#}$The Neural Computing Research
Group, Aston University, Birmingham B4 7ET, UK.}
 \maketitle

\begin{abstract}
The efficacy of a specially constructed Gallager-type error-correcting
code to communication in a Gaussian channel is being examined. The
construction is based on the introduction of complex matrices, used in
both encoding and decoding, which comprise sub-matrices of cascading
connection values. The finite size effects are estimated for comparing
the results to the bounds set by Shannon. The critical noise level
achieved for certain code-rates and infinitely large systems nearly
saturates the bounds set by Shannon even when the connectivity used is
low.  \\
\end{abstract}

\pacs{89.90.+n, 02.50.-r, 05.50.+q, 75.10.Hk}

Information transmission is typically corrupted by noise during
transmission.  Various strategies have been adopted for reducing or
eliminating the noise in the received message. One of the main
approaches is the use of error-correcting codes whereby the original
message is encoded prior to transmission in a manner that enables the
retrieval of the original message from the corrupted transmission.
The maximal transmission rate is bounded by the channel capacity
derived by Shannon~\cite{Shannon} in his ground breaking work of 1948,
which does not provide specific constructions of optimal codes.

Various types of error-correcting codes have been devised over the
years (for a review see~\cite{book}) for improving the transmission
efficiency, most of them are generally still below Shannon's limit. We
will concentrate here on a member of the parity-check codes family
introduced by Gallager~\cite{Gallager}, termed the MN
code~\cite{MacKay} and on a specific construction suggested by us
previously~\cite{us_prl} for the Binary Symmetric Channel (BSC).

The connection between parity-check codes and statistical physics has
been first pointed out in Ref.\cite{Sourlas}, by mapping the decoding
problem onto that of a particular Ising-system with multi-spin
interactions. The corresponding Hamiltonian has been investigated in
both fully-connected\cite{Sourlas} and diluted
systems\cite{ks_sourlas,Saakian} for deriving the typical performance
of these codes; more complex architectures, somewhat similar to those
examined below have been investigated in\cite{kms}, establishing the
connection between statistical physics and Gallager type codes. Most
of these studies have been carried out for a particular channel model,
the BSC, whereby a fraction of the transmitted vector bits is flipped
at random during transmission.

However, different noise models may be considered for simulating
communication in various media. One of the most commonly used noise
models, which is arguably the most suitable one for a wide range of
applications, is that of additive Gaussian noise (usually termed
Additive White Gaussian Noise-AWGN in the literature).  In this
scenario, a message comprising $N$ binary bits is transmitted through
a noisy communication channel; a certain power level is used in
transmitting the information which we will choose to be $\pm 1$ for
simplicity.  The transmitted message is then corrupted by additive
Gaussian noise of zero mean and some variance $\sigma^2$; the received
(real valued) message is then decoded to retrieve the original
message.

The receiver can correct the flipped bits only if the source
transmits $M\!>\!N$ bits; the ratio between the original number of
bits and those of the transmitted message $R\equiv N/M$ constitutes
the code-rate for unbiased messages. The channel capacity in the case
of real-valued transmissions corrupted by Gaussian noise, which
provides the bound on the maximal code rate $R_{c}$, is given
explicitly\cite{Cover} by
\begin{equation}
\label{eq:shannon_bound_real}
R_{c}= \frac{1}{2} \log (1+v^2/\sigma^2) \ ,
\end{equation}
where $v^2$ is the power used for transmission (which we take here to
be $\pm1$) and $v^2/\sigma^2$ is therefore the signal to noise
ratio. However, we will focus here on {\em binary} source messages;
this reduces the maximal code rate to\cite{Cover}
{\small
\begin{equation}
\label{eq:shannon_bound}
R_{c}= \!-\! \int dy P(y) \log P(y) \!+\! \int dy P(y|x=x_0) \log P(y|x=x_0) \ ,
\end{equation}
}
where $x$ is a transmitted bit (of value $x_0\!=\!\pm1$) and $y$ the
received bit after corruption by an additive Gaussian noise, such that
\[
P(y)=\frac{1}{2 \sqrt{2 \pi \sigma^2}} \left[ e^{-(y-x)^{2}/(2 \sigma^2)} +
 e^{-(y+x)^{2}/(2 \sigma^2)} \right] \ .
\]

The specific error-correcting code that we will use here is a
variation of the Gallager code~\cite{Gallager}. It became popular
recently due to the excellent performance of its regular\cite{MacKay},
irregular\cite{Davey,Shokrollahi,Richardson} and the cascading
connection\cite{us_prl} versions.  In the original method, the
transmitted message comprises the original message itself and
additional bits, each of which is derived from the parity of a sum of
certain message-vector bits.  The choice of the message-vector
elements used for generating single code-word bits is carried out
according to a predetermined random set-up and may be represented by a
product of a randomly generated sparse matrix and the message-vector
in a manner explained below. Decoding the received message relies on
iterative probabilistic methods like belief
propagation\cite{MacKay,Frey} or belief revision\cite{Weiss}.

In the MN code one constructs two sparse matrices $A$ and $B$
of dimensionalities $M\!\times\! N$ and $M\! \times\! M$
respectively. The matrix $A$ has $K$ non-zero (unit)
elements per row
and $C (=KM/N)$ per column while $B$ has $L$ per row/column. The matrix
$B^{-1}A$ is then used for encoding the message
\[ {\bf t}_{B} = B^{-1}A \ \bbs \ \mbox{(mod 2)} \ .  \]
The Boolean message vector ${\bf t}_{B}$ is then transmitted as a
vector ${\bf t}$ of {\em real-valued} elements, which we will choose
for simplicity as $\pm1$, and is corrupted by a real-valued noise
vector $\bnu$, where each element is sampled from a Gaussian
distribution of zero mean and variance $\sigma^{2}$. The received
message is of the form
\[ {\bf r} = {\bf t} + \bnu  \ . \]
Using the noise model and the probability of the transmitted bit
being $t_{\mu}=\pm 1$:
\begin{equation}
\label{eq:noise_model}
P(t_{\mu}=\pm1|r_{\mu}) = \frac{e^{-\frac{(t_{\mu}-r_{\mu})^{2}}{2
\sigma^2}}} {e^{-\frac{(t_{\mu}-r_{\mu})^{2}}{2 \sigma^2}}+
{e^{-\frac{(t_{\mu}+r_{\mu})^{2}}{2 \sigma^2}}}} = \frac{1}{ 1+
e^{-\frac{2 t_{\mu} r_{\mu}}{\sigma^2}}} \ ,
\end{equation}
one can easily convert the real-valued noise $\bnu$ to a flip noise
vector such that the probability of an error $n_{\mu}=1$ (error) is
\begin{equation}
\label{eq:error_prob}
P(n_{\mu}=1) = \frac{1}{ 1+ e^{-\frac{2 r_{\mu}}{\sigma^2}}} \ .
\end{equation}
Note that $P(n_{\mu}=1)$ may be larger than $1/2$.  The noise vector
$\bn$ and our estimate for the transmitted vector $\hbt$ are defined
probabilistically by using the probabilities derived in
Eq.(\ref{eq:error_prob}) and Eq.(\ref{eq:noise_model}) respectively.

Having an estimate for the transmitted vector $\hbt$ as well as an
estimate for the noise vector $\bn$, one decodes the binary
received message $\hbt$ by employing the matrix $B$ to obtain:
\begin{equation}
\label{eq:BP_equation}
{\bf z} = B  \ \hbt = A \bbs + B {\bf n} \ .
\end{equation}
This requires solving the equation
\[ [A , B] \left[ \begin{array}{c}
{\bbs '} \\ \bn '
\end{array} \right] =  {\bf z}  \ , \]
where $\bbs '$ and $ \bn '$ are the unknowns. This is being carried out
here using methods of belief network decoding\cite{MacKay,Frey}, where
pseudo-posterior probabilities, for the decoded message bits being 0
or 1, are calculated by solving iteratively a set of equations for the
conditional probabilities of the codeword bits given the decoded
message and vice versa. For exact details of the method used and the
equation themselves see\cite{MacKay}. Two differences from the
framework used in the case of a Binary Symmetric Channel (BSC) that
should be noticed: 1) The probabilities of Eq.(\ref{eq:error_prob})
and Eq.(\ref{eq:noise_model}) may be used for defining the priors for
{\em single components} of the noise and signal vectors respectively. 2)
Initial conditions for the noise part of the dynamics may also be
derived using Eq.(\ref{eq:error_prob}).

The key point in obtaining improved performance is the construction of
the matrices $A$ and $B$. The original MN code\cite{MacKay} as well as
that of Gallager\cite{Gallager} advocated the use of regular
architectures with fixed column connectivity; it also suggested that
fixed $K$ values may be preferred. Recent work in the area of
irregular codes~\cite{Davey,Shokrollahi,Richardson} suggest that
irregular codes have the potential of providing superior performance
which nearly saturates Shannon's limit.  These methods concentrate on
different column connectivities and use high $K$ and $C$ values (up to
50), which of course increase the complexity of the algorithm and the
decoding time required. Decoding delays are of major consideration in
most practical applications.

Our method uses the same structure as the MN codes and builds on
insight gained from the study of physical systems with symmetric and
asymmetric\cite{ido} multi-spin interactions and from examining
special cases of Gallager's method\cite{ks_sourlas,kms}.  Our previous
studies for the binary symmetric channel\cite{us_prl} suggest that a
careful construction, based on different $K$ and $L$ values for the
sub-matrices of $A$ and $B$ respectively, while keeping the
connectivity of each of the sub-matrices (and of the matrix as a
whole) as uniform as possible, will provide the best results.  The
guidelines for this architecture are given below and come from the
mean-field calculations of Refs.\cite{us_prl,us_pre}, showing that the
choice of low $K$ and $L$ value codes results in a large basin of
attraction but imperfect end-magnetisation, while codes with higher
$K$ and $L$ values can potentially saturate Shannon's bound but suffer
from a rapidly decreasing basin of attraction as $K$ and $L$ increase.
To exploit the advantages of both architectures and obtain optimal
performance, a cascading method was suggested\cite{us_prl,us_pre}
whereby one constructs the matrices $A$ and $B$ from sub-matrices of
different $K$ and $L$ values; such that lower values will drive the
overlap increase between the decoded and the original messages to a
level that enables the higher connectivity sub-matrices to come into
play, allowing the system to converge to the perfectly decoded
message\cite{us_pre}.

Optimising the trade-off between having a large basin of attraction
and improved end magnetisation can be done
straightforwardly\cite{us_pre} in the case of simple codes
\cite{Sourlas} but is not very easy in general. Guidelines for
optimising the construction in the general case have been provided in
Ref.\cite{us_prl}; the key points include: 1) The first sub-matrices
are characterised by low $K$ and $L$ values ($\le 2$), while $K$
values in subsequent sub-matrices are chosen gradually higher, so as
to support the correction of faulty bits, and $L=1$.  2) Keeping the
number of non-zero column elements as uniform as possible (preferably
fixed).  3) To guarantee the inversion of the matrix $B$, and since
noise bits have no explicit correlation, we use a patterned structure,
$B_{i,k} \! =\!  \delta_{i,k} \! +\!  \delta_{i,k+5}$, for the
$B$-sub-matrices with $L\!  =\!2$ and $B_{i,k} \! =\! \delta_{i,k}$ for
$L\! =\!1$.  4) The sub-matrix with the lowest $K$ value, which
dominated the dynamics in the initial stage, low magnetisation, has to
include some odd $K$ values in order to break the inversion symmetry,
otherwise the two solutions with $m\! =\!\pm1$ are equally attractive. It
was also found to dramatically improve the convergence times.

We will now focus on two specific architectures, constructed for the
cases of $R=1/2$ and $R=1/4$, for demonstrating the exceptional
performance obtained by employing this method. In each of the cases we
divided the composed matrix $[A | B]$ to several sub-matrices
characterised by specific $K$ and $L$ values as explained in table 1;
the dimensionalities of the full $A$ and $B$ matrices are $M
\!\times\! N$ and $M \!\times\! M$ respectively. Sub-matrix elements
were chosen at random (in matrix $A$) according to the guidelines
mentioned above. Encoding was carried out straightforwardly by using
the matrix $B^{-1}A$. The corrupted messages were decoded using the
set of recursive equations of Ref.\cite{MacKay}, using random initial
conditions for the signal while the initial conditions for the noise
vector where obtained according to the noise and signal probabilities 
Eq.(\ref{eq:error_prob}).  The prior probabilities of 
were chosen according to Eqs.(\ref{eq:error_prob}) and
(\ref{eq:noise_model}).

In each experiment, $T$ blocks of $N$-bit unbiased messages were sent
through a Gaussian noisy channel of zero mean and variance
$\sigma^{2}$ (enforced exactly); the bit error-rate, denoted $p_{b}$,
was monitored.  We performed between $T\!=\!10^4 - 5 \times 10^4$ trial runs
for each system size and noise level, starting from different initial
conditions. These were averaged to obtain the mean bit error-rate and
the corresponding variance.  In most of our experiments we observed
convergence after less than 100 iterations, except very close to the
critical noise level.  The main halting criterion we adopted relies on
either obtaining a solution to Eq.(\ref{eq:BP_equation}) or by the
stationarity of the first $N$ bits (i.e., the decoded message) over a
certain number of iterations.  One should also mention that the
decoding algorithm's complexity is of $O(N)$ as all matrices are
sparse. The inversion of the matrix $B$ is carried out only once and
requires $O(1)$ operations due to the  structure chosen.

The construction used for the matrices in these two cases appear in
table 1 as well as the maximal standard deviation $\sigma^{N}_{c}$ for
which $P_b< 2 \times 10^{-5}$ for a given message length $N$, the
predicted maximal standard deviation $\sigma^{\infty}_{c}$ once finite
size effects have been considered (discussed below) and Shannon's
maximal standard deviation $\sigma_{c}$ defined in
Eq.(\ref{eq:shannon_bound}).  These results, as well as other results
reported here, could be improved upon by avoiding matrices with small
loops and by replacing the method of belief propagation by belief
revision (our random construction of the matrix $A$ even allows for
small loops of size one).  It was shown that both improvements have a
significant impact on the performance of this type of
codes\cite{MacKay,Weiss}.  With these improvements, the actual bit
errors is expected to be typically lower than the reported value of
$P_b=2 \times 10^{-5}$; however, as we have been limited to about $T=5
\times 10^4$ trials per noise value we can only provide an upper bound
to the actual error values.

To compare our results to those obtained by using turbo
codes\cite{turbo} and in Ref.\cite{Richardson} we plotted in Fig.1 the
two curves (dotted and dashed respectively), for $N=10^3$ and $10^4$,
against the results obtained using our cascading connection method
(filled triangles). It is clear from the figure that results obtained
using our method are superior in all cases examined.  Furthermore,
from table 1, one can conclude that the averaged connectivity,
$\overline{C}$ in the case of $R=1/2$ and $1/4$ is $5$ and $9$
respectively for the matrix $A$ and $3/2$ for the matrix
$B$. Similarly, the averaged $K$ values for $R=1/2$ and $1/4$ are
$\overline{K}=5/2$ and $9/4$, respectively. These number are much
smaller than those used in Refs.\cite{Shokrollahi,Richardson} and
other irregular constructions.  Minimising $\overline{K}$ and
$\overline{C}$ is of great interest to practitioners since decoding
delays are directly proportional to the $\overline{K}$ and
$\overline{C}$ values used\cite{MacKay}.

It is clear from Fig.1 that the finite size effects are significant in
defining the code's performance. It is therefore desirable to find the
performance in the limit of infinite messages which are also assumed
in deriving Shannon's bound. We employ two main methods for studying
the finite size effects: a) The transition from
perfect ($m(\sigma)\!=\!1$) to no retrieval ($m(\sigma)\!=\!0$), as a
function of the standard deviation $\sigma$, is expected to become a
step function (at $\sigma^{\infty}_{c}$) as $N\!\rightarrow\! \infty$;
therefore, if the percentage of perfectly retrieved blocks in the
sample, for a given standard deviation $\sigma$, increases (decreases)
with $N$ one can deduce that $\sigma\! <\!  \sigma^{\infty}_{c}$ (or
$\sigma\! >\!  \sigma^{\infty}_{c}$). b) Convergence times near
criticality usually diverge as $1/(\sigma^{\infty}_{c}-\sigma)$; by
monitoring average convergence times for various $\sigma$ values and
extrapolating one may deduce the corresponding critical standard
deviation.

Both methods have been used in finding the critical values for $R=1/2$
and $R=1/4$; the results obtained appear in table~1.  In Fig.2 we
demonstrate the two methods: we ordered the samples obtained for
$R\!=\!1/2$, $\sigma\!=\!0.915,0.935$ (dashed and solid lines
respectively) and $N\!=\!1000,10000$ (thin and thick lines
respectively) according to their magnetisation; results with higher
magnetisation appear on the left and the $x$ axis was normalised to
represent fractions of the complete set of trials. One can easily see
that the fraction of perfectly retrieved blocks increases with system
size indicating that $\sigma < \sigma^{\infty}_{c}$. In the inset one
finds log-log plots of the mean convergence times $\tau$ for
$R\!=\!1/2,1/4$ and $N=10000$ carried out on perfectly retrieved
blocks with less than 3 error bits. The optimal fitting of expressions
of the form $\tau \propto 1/(\sigma^{\infty}_{c}-\sigma)$ provides
another indication for the $\sigma^{\infty}_{c}$ values, which are
consistent with those obtained by the first method.

We end this presentation by discussing the main difference between our
method and those presented in
Refs.\cite{Davey,Shokrollahi,Richardson}.  Firstly, our construction
builds on sub-matrices of different $K$ and $L$ values keeping the
connectivity in each of the columns as uniform as possible; this
equates the corrections received by the various bits while allowing
them to participate in different multi-spin interactions, so as to
provide contributions of different types throughout the dynamics. In
contrast, other irregular codes build on the use of different column
connectivities such that a small number of bits, of high connectivity,
will lead the decoding process, gathering more corrected bits as the
decoding progresses. Secondly,
Refs.\cite{Davey,Shokrollahi,Richardson} as well as others point to
the need of high multi-spin interactions for achieving performance
close to Shannon's bound; we show here that low $K$, $L$ and $C$
values are sufficient for near-optimal performance (in the case of
$R=1/2$ and $1/4$ the averaged connectivities are $\overline{C}=5$ and
$9$ respectively for the matrix $A$ and $3/2$ for the matrix $B$),
allowing one to carry out the encoding and decoding tasks
significantly faster.  Our work suggests that it is possible to come
very close to saturating Shannon's bound with finite connectivity, at
least for the code rates considered here.  It is plausible that operating
close to $R\!=\!1$ will require higher $K$, $L$ values and may require
infinite $C$ or $\overline{C}$ values; this question is currently
under investigation.

We have shown that through a successive change in the number of
multi-spin interactions ($K$ and $L$) one can boost the performance of
Gallager-type error-correcting codes.  The results obtained here for
the case of additive Gaussian noise suggests competitive performance
to similar state-of-the-art codes for finite $N$ values; extending the
results to the case of infinitely large systems suggest that the
current code is less than $0.1$dB from saturating the theoretical
bounds set by Shannon.  It would be interesting to examine methods for
improving the finite size behaviour of this type of codes; these would
be of great interest to practitioners.

\vspace*{3mm}

{\small
\vspace*{0.1in} {\bf \hspace*{-1em} Acknowledgement} We would like to
thank Dr.~Yoshiyuki Kabashima for helpful discussions.}

\vspace*{3mm}
%%%%%%%%%%%%%%%%%%%%%%%%%%%%%%%%%%%%%%%%%%%%%%%%%%%%%%%%%%%%%%%%%%%%%%
%\newpage

\onecolumn

\begin{table*}
\begin{tabular}{|c|c|c|c|c|c|c|c|} 
R &  $A$ & $K$ & $B$ & $L$ &  $\sigma^{10000}_{c}$/(dB) &
$\sigma^{\infty}_{c}$/(dB) & $\sigma_{c}$/(dB) \\ \hline  \hline
$1/2$ & $1/10 \ N\!\times\! N$ & 1 & $1/10 \ N\!\times\! 2N$ & 2 & 0.89 &
0.973  & 0.979   \\
  &  $9/10 \ N\!\times\! N$ & 2 & $9/10 \ N\!\times\! 2N$ & 2 &  (1.012) &
(0.238) & (0.185) \\
  &  $3/4 \ N\!\times\! N$ & 2 & $3/4 \ N\!\times\! 2N$ & 1  &  & &  \\
  &  $3/20 \ N\!\times\! N$ & 6 & $3/20 \ N\!\times\! 2N$ & 1 &   & &  \\
  &  $1/10 \ N\!\times\! N$ & 7 & $1/10 \ N\!\times\! 2N$ & 1 &  & &  \\
 \hline
$1/4$  & $3/2 \ N\!\times\! N$ & 1 & $3/2 \ N\!\times\! 4N$ & 2
&  1.45  & 1.537 &  1.550 \\
  &  $N/2 \!\times\! N$ & 4 & $N/2 \!\times\! 4N$ & 2 &
(-0.217) & (-0.721) & (-0.797) \\
  &  $1/3 \ N \!\times\! N$ & 4 & $1/3 \ N \!\times\! 4N$ & 1 &  & & \\
   &  $5/6 \ N \!\times\! N$ & 3 & $5/6 \ N \!\times\! 4N$ & 1 &  & &  \\
   &  $ 5/6 \ N \!\times\! N$ & 2 & $ 5/6 \ N \!\times\! 4N$ & 1 &  & &  \\
\end{tabular}

\vspace*{1cm}
\caption{The critical noise standard deviation $\sigma^{N}_{c}$ and
$\sigma^{\infty}_{c}$ obtained by employing our method for various
code rates in comparison to the maximal standard deviation
$\sigma_{c}$ provided by Shannon's bound. Details of the specific
architectures used and their row/column connectivities are also
provided.}

\end{table*}

\twocolumn
\begin{figure}
\label{fig:1}
\begin{center}
\epsfysize = 6.0cm
\epsfbox[0 120 670 520]{GR5_Pb.eps}
%\epsfysize = 10.0cm
%\epsfbox[-50 120 670 520]{../../figures/Pb_f.eps}
\end{center}
\vspace*{2cm}
\caption{Bit-error rate $p_{b}$ as a function of the standard
deviation for a given code-rate $R\!=\!1/2$ for systems of size
$N\!=\!1000,10000$ (right and left respectively). Our results for each
system size appear as black triangles, while results obtained via the
turbo code and in Ref.[13] for systems of similar sizes
appear as curves (dotted and dashed respectively). }
\end{figure}

\newpage

\begin{figure}
\label{fig:2}
\begin{center}
\vspace*{-1cm}
\epsfysize = 6.0cm
\epsfbox[0 120 670 520]{GR5_profile.eps}
\epsfysize = 4.4cm \epsfbox[-130 -415 420 80]{GN1_4R525_convtimes.eps}
%\epsfysize = 10.0cm
%\epsfbox[0 120 670 520]{../../figures/GR5_profile.eps}
%\epsfysize = 7.8cm \epsfbox[-130 -415 42080]{new.eps}
\end{center}
\vspace*{-3cm}
%\vspace*{-1cm}
\caption{The block magnetisations profile for $R=1/2$,
$\sigma=0.915,935$ (dashed and solid lines respectively) and
$N=1000,10000$ (thin and thick lines respectively), showing the sample
magnetisation $m$ vs. the fraction of the complete set of trials. A
total of about 10000 trials were rearranged in a descending order
according to their magnetisation values.  One can see that the
fraction of perfectly retrieved blocks increases with system
size. Inset - log-log plots of mean convergence times $\tau$ for
$N=10000$ and $R=1/2,1/4$ (white and black triangles
respectively). The $\sigma^{\infty}_{c}$ values were calculated by
fitting expressions of the form $\tau \propto
1/(\sigma^{\infty}_{c}-\sigma)$ through the data.}
\end{figure}

\end{document}